# The influence of Bi$_2$O$_3$ glass powder in silver paste on the fabrication of ohmic contacts and its potential effectiveness in solar cells and integrated circuits on p-type silicon substrates

Jung-Ting Tsai[1*], Li-Kai Lin[1], Shun-Tian Lin[3], Lia Stanciu[1], Martin Byung-Guk Jun[2]

[1]*School of Materials Engineering, Purdue University, West Lafayette, IN 47907, USA*

[2]*School of Mechanical Engineering, Purdue University, West Lafayette, IN 47907, USA*

[3]*Department of Mechanical Engineering, National Taiwan University of Science and Technology, Taipei 106, Taiwan*

\* Email of Corresponding Author: tsai92@purdue.edu (J.-T. Tsai)

## Abstract

The present work critically investigates the influence of low-melting glasses on the fabrication of metal contacts, with the goal of advancing applications of bismuth-based oxide glass and screen-printed silver contacts for use in integrated circuits (ICs), solar cells, and sensors. In this study, novel electrode contacts were fabricated by screen printing composite pastes composed of mainly silver powder, Bi$_2$O$_3$ glass powder, and acyclic binder, and then firing the pastes in a belt furnace. The microstructures of the composite films after firing at 830~890°C were observed under different corrosion conditions, and the resulting layers were analyzed with X-ray diffraction (XRD) and the transfer length method (TLM). A series of investigations to determine the influence of Bi$_2$O$_3$ glass in silver paste involved various tests, including differential thermal analysis (DTA), scanning electron microscopy (SEM), electron probe X-ray microanalysis (EPMA), secondary ion mass spectrometry (SIMS) and transmission electron microscopy (TEM), to determine the effects of Bi$_2$O$_3$ mixed with silver and the efficacy of the resulting metal contacts in IC fabrications. It was observed that the additive, Bi$_2$O$_3$ glass, controlled the melting of the silver into the glass, influencing the precipitation of Ag crystallites. In addition, an increase in the firing temperature caused excessive growth of Ag crystallites and current leakage, and the size and the relationship of the Ag crystallites in the Bi$_2$O$_3$ glass were confirmed.

# Introduction

In integrated circuits (ICs), solar cells, and various electronics, the most important material is the conducting paste, which facilitates the formation of metal contacts. The electrode contacts on top of the source, drain, and gate regions of a transistor or the front-side grid of a solar cell are usually fabricated by screen printing silver paste and then firing in a belt furnace. The composition, content, proportion, and process parameters of a conducting paste can influence the performance of the final electrode product. The existing commercial Ag pastes on the market consist mostly of Ag powder, lead-glass frit, and an organic transport system [7]. The concentration of glass frit is usually less than 5 wt%. The melting characteristics of the glass frit and the dissolved Ag significantly influence the contact resistance and fill factors (FFs). In addition, the glass frit forms a glass layer between the bulk Si and Ag, which can further react with the bulk Si to precipitate Ag on the Si surface upon high-temperature firing [6, 8]. Furthermore, the glass frit plays a critical role in the contact quality because it can affect the adhesion, conduction, and contact formation between layers.

Of the main materials used to fabricate conducting paste, glass frit greatly improves the conductivity nd alters the microstructure of the paste. In recent decades, lead oxide (PbO)-based glass systems have become popular due to their high structural stability, low glass transition temperature ($T_g$) and good thermal and electrical characteristics. However, recent environmental regulations have restricted the wide use of PbO systems, so the development of Pb-free or low-Pb materials, which can replace pure PbO, has been of interest. Since bismuth oxide ($Bi_2O_3$)-based glasses are thermally stable against phase transformation, $Bi_2O_3$ has been investigated as a candidate material to replace PbO [9]. When $Bi_2O_3$ is used, a large amount of glass network modifier is necessary to obtain low glass transition temperatures, unlike in a PbO glass system [10, 11]. In addition, low-melting $Bi_2O_3$ glass has been widely used in various electronics, sensors, and thick-film (TF) technologies, such as for mechanical attachment in sensors, semiconductor encapsulation, and hermetic sealing of sensors. As a bulk material, low-melting glass can be found in crystal glass, cathode ray tubes (CRTs), optical devices, and γ-ray shields. In layered structures, $Bi_2O_3$ glass has been evaluated for hermetic package sealing, sensor sealing, TF overglazes, TF conductors, and TF resistors.

Many studies have shown that $Bi_2O_3$ glass is crucial in many fields, but its application is not limited to the use of $Bi_2O_3$ alone. In fact, numerous applications of $Bi_2O_3$ as an additive in ohmic contacts have revealed it high value, but many aspects of the material properties of silver paste containing additives remain unknown. Thus, more comprehensive, specific, and sound experimental results for the application of $Bi_2O_3$ glass with silver paste and the associated material properties are urgently needed [40].

During the fabrication of metal contacts, the main ways to improve the performance of ohmic contacts, improve the front-side grid, and increase the integrity of electrodes are to (i) decrease the finger-line resistance, (ii) decrease the finger-line width, (iii) increase the aspect ratio, and (iv) reduce the contact resistivity of the low-doped emitters [1, 2]. In addition, during the fabrication process, the substrate can have significant effects on the metal structure above it. Ag–Si surface contacts have been widely studied, but there are still interesting intrinsic aspects of their microstructural changes, material interactions, and conductivity changes throughout the fabrication process that are not fully understood. Comprehensive investigation of these aspects is essential to improving the efficiency of electronics. Furthermore, several studies have reported that a good-quality ohmic contact is controlled by the properties of Ag crystallites and the glass layer thickness; thus, the firing process and glass frit are key factors for improving the efficiency of electronics [3-6]. In this study, the properties of the low-resistivity and highly doped contacts in pastes are investigated in homemade chips to better understand the nature behind their behavior.

$Bi_2O_3$-based glass powders were prepared with different formulas and compositions ($Bi_2O_3$, $B_2O_3$, and $P_2O_5$), which resulted in different properties and stabilities from those of commercial $Bi_2O_3$ glass powder ($Bi_2O_3$ and $B_2O_3$). It is of fundamental and technological interest to examine whether this homemade $Bi_2O_3$-based glass frit containing an additive plays a role similar to that of commercial glass powder in the formation of Ag crystallites. The effects of different glass transition temperatures and the silver electrode morphology were also investigated to improve the performance of the resulting ohmic contacts on both solar cells and ICs (Fig. 1).

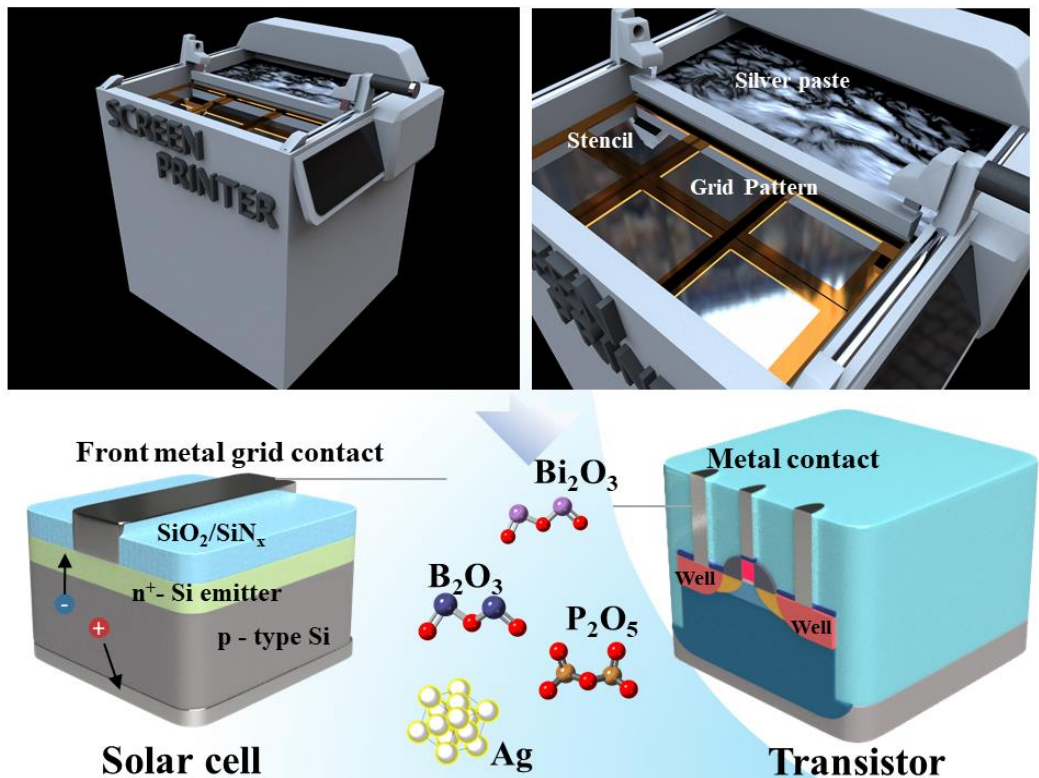

Fig. 1 Illustration of the screen printing process. (Top-Left) Overview of the screen printer. (Top-Right) Focused view of the operation panel. The application of the metal composite on (Bottom-Left) a solar cell and (Bottom-Right) a transistor.

**Experimental Procedures**

The development of the metal pastes in this study is based on several important factors: (1) the shape and the proportion of metal particles, (2) the content of organic carriers, and (3) the thermal properties of the glass powder. The developed pastes contain three major components. The first is commercial silver powder with spherical particles. The amount of silver particles was controlled at 80~85 wt%. The second is the commercial polymer, or organic carrier, which was 10~15 wt%. The final component is the glass powder, which was added to a content of 3~5 wt%.

To prepare the glass powder, high-purity $Bi_2O_3$ (Sigma-Aldrich Co., USA) (MAIDO Co., Ltd, 99.9%, Japan), $B_2O_3$ (MAIDO Co. Ltd., 95.3%, Japan), and $P_2O_5$ (Hoechst Co. Ltd., 99.7%, Japan) were used as raw starting materials. The powder was separated into 10 g quantities, which were placed in small plastic bottles, dry mixed by a ball mill without steel balls for 30 min, and then placed in a 50 ml alumina pot. The pot was placed in a 1050~1150°C air furnace, which heated the molten materials into a liquid by maintaining the temperature for 30 min. The liquid glass was poured into a water tank for rapid cooling. The glass powder was placed in a plastic bottle and then milled with zirconium balls for 14 days. The thermal properties ($T_g$) of the glass frits were determined in the beginning by thermogravimetric-differential thermal analysis (TG-DTA; TG8120 Rigaku) as follows: The glass powder (20~30 mg) was placed in a DTA instrument (Seiko SSC5000), and the temperature was increased from room temperature to 800°C at 10°C/min, with air as the carrier gas (flow rate of 100 cc/min). The particle sizes of the commercial and homemade glass were measured by laser particle size analysis (Mastersizer 2000E). The absence of crystalline phases was then confirmed by XRD (Bruker Co., USA) (Model-D/MaxIIB, Riau, Japan). The silver powder, glass powder, and acrylic binder (butyl carbitol, DB, terpinel, α-terpineol, β-terpineol, texanol, diethylene glycol monobutyl ether acetate, cellulose, nitrocellulose, and polymethylmethacrylate) were then mixed in a weight ratio of 80:3:17, placed in a 3-roll mixer, and homogenized until the high-molecular-weight polymer was monodispersed in the colloid. After the pastes were prepared, they were printed on silicon substrates, and the substrates were placed in a dryer for 30 min. When the solvent in the paste had evaporated, the substrates were removed for final sintering at 830~890°C.

The microstructures of the metal contact were examined by corrosion tests. The degree of corrosion on the surface was observed at different stages. First, the surface of the silver electrode was corroded by nitric acid (99.8% HCl + 99.6% $H_2O_2$, 1:1, 10 min), hydrofluoric acid (4% HF, 2 min), and nitric acid (99.8% HCl + 99.6% $H_2O_2$, 2:8, 10 min). Four different surfaces can be observed after each corrosion stage. The relationships between each set of layers are discussed for the following four stages: (a) when the surface was completely covered by silver paste (uncorroded); (b) when the glass layer covered the crystallized silver; (c) when crystallized silver particles had precipitated on the silicon wafer surface; and (d) for different residual trace holes on the silicon wafer surface (silicon substrate). The microstructures on the surface were examined using scanning electron microscopy (SEM; JEOL-6500F) with energy dispersive spectroscopy (EDS) on an instrument operating at an accelerating voltage of 20 kV. X-ray diffraction (XRD, D2 Phase) was used for phase analysis of the silver surface. The specific contact resistance was determined by using the transfer length method (TLM) pattern, which was printed at the same time as the pattern on silicon.

The samples were then cut into pieces to reveal the cross-sectional microstructure using SEM (FEI Corp., Quanta, USA) (JEOL, JSM-6060, Japan). Electron probe X-ray microanalysis (EPMA; JEOL, JXA-8600SX, Japan) was used to observe the fractured section of the silver paste/wafer material. The accelerating voltage was 20 kV, and a wavelength dispersive spectrometer (WDS) was used to quantitatively map the silver, oxygen, bismuth, silicon, and phosphor. Secondary ion mass spectrometry (SIMS; Cameca, IMS-4f, France) was used to determine the composition below the corroded wafer surface, as the secondary ions detected were cations, which were from the silver and bismuth at a depth of 2 μm. In this experiment, $O^{2+}$ was selected as the ion source. Transmission electron microscopy (TEM) (FEI Corp., Tecnai, USA) (Philips FEI-TEM) was used to study the microstructures and features at the glass interface.

## Results and Discussion

Before fabrication, the particle sizes of the commercial and homemade glass and silver particles were measured by laser particle size analysis and SEM to be 8.81 μm, 1.18 μm, and 1.24 μm, respectively, as shown in Fig. 2.

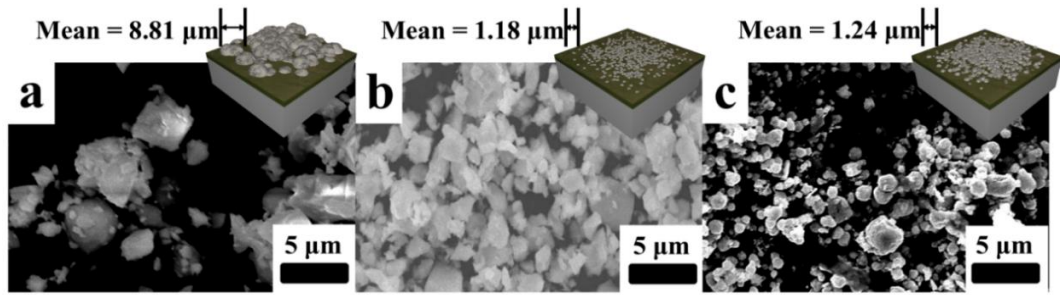

Fig. 2 Morphology of the (a) commercial glass (CG), (b) homemade glass (BPB) and (c) silver (Ag) powders used in this study.

After the metal contacts were prepared, their microstructures were examined by layer-by-layer corrosion tests. The surfaces of different layers were observed by four different degrees of corrosion applied at different stages as follows:

In stage one, the samples fired from 860°C~890°C had no pores. The silver grains were fully wetted by the glass, and the glass phase was dispersed between the grains. The wetting of the glass phase was better than that of the sample fired at 830°C; however, there was coarsening of the surface, as shown in Fig. 3 A (a)~(c).

In the second stage of corrosion, contact between the crystallized silver and the substrate is difficult to determine from the SEM images. As shown in Fig. 3 A (d), the contrast of the large precipitated particles was detected by EDS with SEM (region 2), as shown in Fig. 3 B. The composition analysis results show that this region contained silver, silicon, bismuth, and phosphorous (in order of intensity), as shown in Fig. 3 A (a). The contrast of the surface without precipitates (region 3) that contained silicon, silver, bismuth, aluminum, and oxygen (in order of intensity) is shown in Fig. 3 A (b). It is noteworthy that although the two regions have silver and silicon signals, there is an obvious intensity difference. Region 2 has an additional oxygen signal, indicating that the Ag melted in the glass was crystallized and precipitated. Ag particles that were not fully precipitated melted into the glass layer again, which is why the Ag signal occurred when using EDS even when there were no Ag precipitates [6, 12]. It is obvious that the glass did not disperse uniformly on the surface but accumulated in separate regions. In addition, there were many small circular particles in these regions that were also concentrated under the glass layer. Silver, silicon, bismuth, and phosphor signals were detected, indicating that the spherical silver nanoparticles were encased in a glass layer, as shown in Fig. 3 A (e) ~ (f) by the red arrows.

The particle size and distribution of the Ag crystallites on the silicon wafer surface can be clearly observed during the third stage, as shown in Fig. 3 A (g) ~ (i). The purpose during this stage was to corrode the silicon oxide and original bismuth oxide glass layer. Ag crystallites of different sizes contacted the silicon substrate. As measured by the scale marker, the largest particle was 1 μm in diameter, and the smallest particle was 0.1 μm in diameter. Fig. 3 A (g) shows that some Ag precipitates left a square concave depression in the surface after corrosion by hydrofluoric acid, and these depressions became very obvious as the temperature increased.

In the fourth stage, on the basis of the microstructure, the corroded Ag crystallites left complete traces on the silicon wafer surface. Different traces were left on the substrate when the Ag crystallites grew in different directions on the silicon wafer surface. As seen in Fig. 3 A (j), the traces left by the Ag crystallites were spread over the entire surface, and the pores were upended polyhedrons; moreover, there were upended rectangular quadrilateral prisms. In Fig. 3 A (k), the pore shape tended to be large, and the pores were almost polyhedrons. In Fig. 3 A (l), it is obvious that the pore shape and depth had increased; thus, some quadrilateral prisms were deformed.

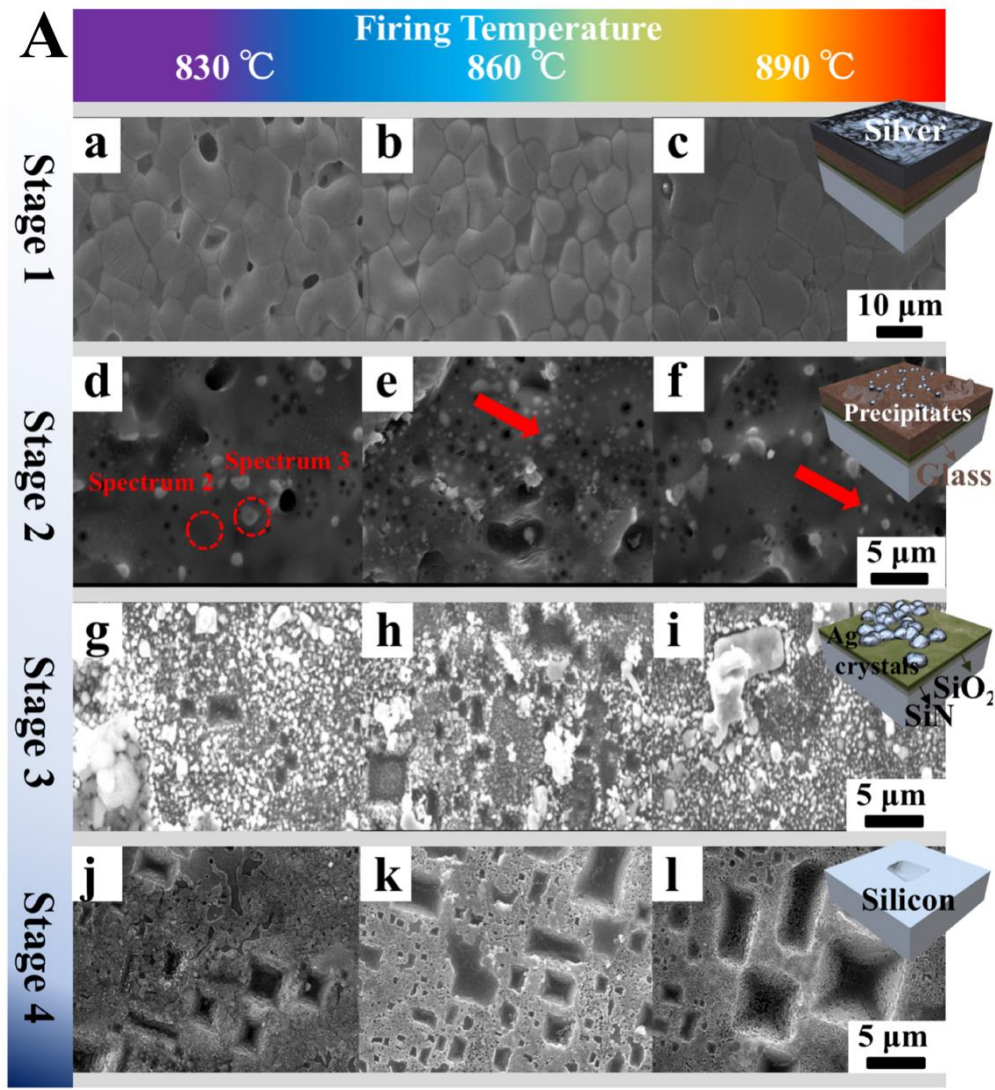

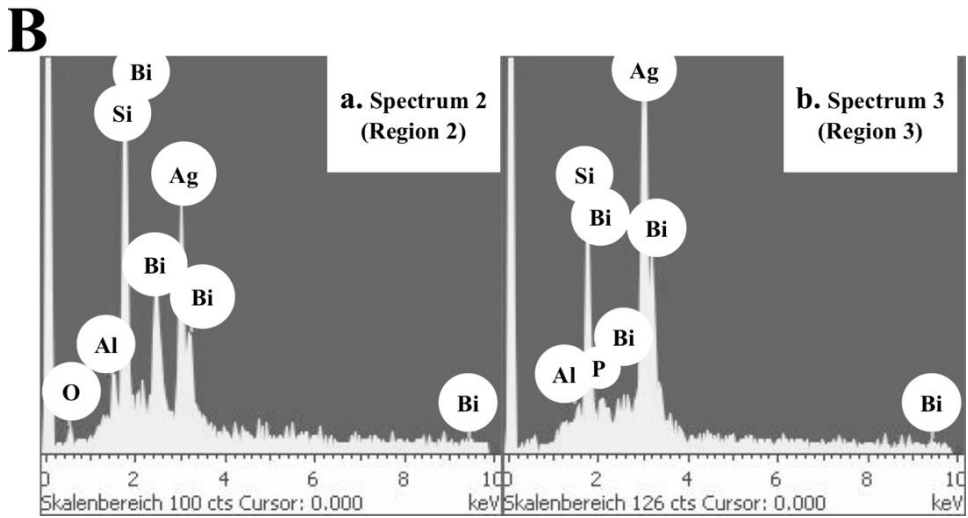

Fig. 2 (A) Microstructures of silver paste fired at 830~890°C in different corrosion stages. (B) EDS analysis of regions 2 and 3 of the silver paste immersed in HNO₃ solution in A. (d) SEM micrograph.

XRD was used to confirm that the crystalline phases were present before and after the corrosion stages. Fig. 4 A shows the XRD pattern after the silver paste was sintered from 830°C~890°C. Only the crystalline phase of Ag existed; no glass crystalline phase could be detected, which may be related to the addition of a small amount of glass powder. Moreover, there was no indication of oxidation of the silver paste on the sintered surface. The (111) peak was suppressed, and the (200) peak value was relatively strong. Thus, the silver had a preferred (200) orientation during sintering [1, 2].

Fig. 4 B shows the XRD pattern after the silver paste was corroded. The Ag layer was corroded while the substrate and upper glass layer were maintained. Fig. 4 B shows that the main (111) Ag peak was followed by secondary (200) (220) peaks. This result is obviously different from that in Fig. 4 A. The background XRD spectrum of the corroded substrate contained some minor background signals from 20°~35°, which were related to the large amount of molten silver powder in the glass. The obtained data fit the result that Karski and Witonska obtained [13]. In particular, there was no signal from the silicon substrate, which corresponds to the rapid heating from 830~890°C; thus, the silicon melted at the interface because the eutectic point for Ag and silicon is 835°C, whereas the eutectic point for Ag and glass ($Bi_2O_3$) is only 687°C [14]. XRD revealed the presence of crystalline Ag, while the silicon and glass amorphous phases were detected as background signals [15]. Previous studies indicated that when an insulator is used as a substrate, the valence electrons prevent the growth of monocrystals, which explains why only Ag signals were detected on the silicon substrate [16].

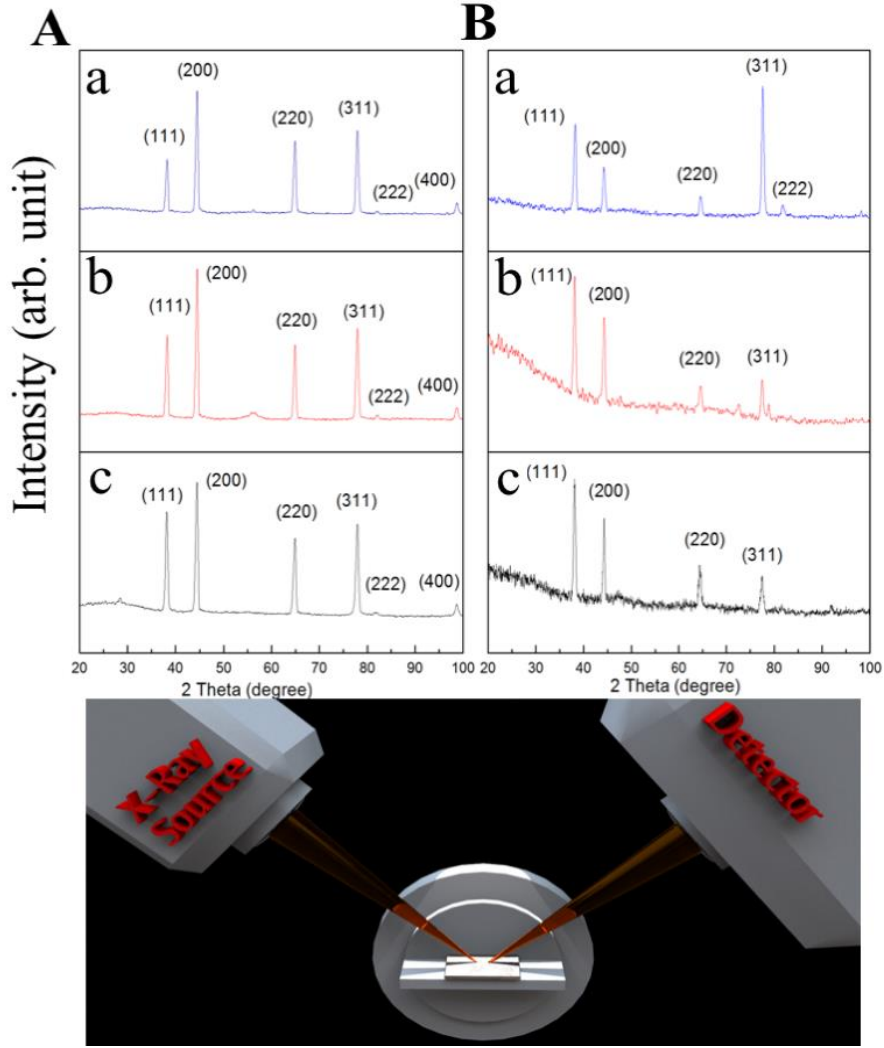

Fig. 4 XRD pattern (A) before corrosion and (B) after corrosion of silver paste fired at (a) 830℃, (b) 860℃ and (c) 890℃, respectively.

Multiple studies have proven that the addition of a small amount of glass improves the mechanical and electrical properties of the Ag layer and silicon substrate during the early stages [3-5]. Schubert and Huster [17] have developed many hypotheses regarding the contact resistance of Ag crystallites, and the contact resistance of the Ag layer and silicon substrate was measured with the TLM. It is believed that the relationship between the contact resistance and the amount of Ag crystallites is key to the contact resistance. If the contact resistance of individual Ag crystals at the interface is known, then the contact resistance can be effectively controlled, and the efficiency can be increased. Therefore, the contact resistance of individual Ag crystals was measured using the method proposed by Kontermann and Preu [18], which can be expressed as follows:

$$\frac{\rho_{c,crystal}}{A_{crystalline\ silver} \cdot A_{crystalline\ silver,direct}} = \rho_C \quad (1)$$

where $\rho_{c,\ crystal}$ is the silver crystal found at the interface of the silver thick film, $A_{crystalline\ silver}$ is the fraction of surface area covered by crystalline silver, A$_{crystalline\ silver,\ direct}$ is the fraction of crystalline silver in direct contact with the silver bulk, and $\rho_c$ is the specific contact resistance of the silver thick film.

First, from 830~890°C, the special contact resistance was 1.918x10$^{-4}$ Ω.cm$^2$, 2.922x10$^{-4}$ Ω.cm$^2$, and 6.322x10$^{-4}$ Ω.cm$^2$, and the Ag crystallite coverage on the entire surface obtained from all measured diameters according to Fig. 3 A (j)~(l) was 15%, 25% and 35%. The ratio of crystallites contacting the silver paste layer was 0.1%, and this value was substituted into Eq. (1) to obtain the value for $\rho_{c,\ crystal}$. The resistance of individual Ag crystals from 830~890°C was 2.9x10$^{-8}$ Ω.cm$^2$, 7.3 x10-8 Ω.cm$^2$, and 2.2 x10-7 Ω.cm$^2$, as shown in Table 1.

Schubert and Huster [17] also calculated the contact resistance to be 2x10$^{-8}$ Ω.cm$^2$, which is close to the value calculated herein (2.2x10$^{-8}$~7.3x10$^{-8}$ Ω.cm$^2$). Therefore, we believe that the resistance in individual silver crystals can reach 10$^{-8}$ Ω.cm$^2$.

After we evaluated the effects of Ag crystallites, the influence of the silver powder on the fabrication of metal contacts was investigated. First, the conductivity tended to increase as the amount of silver powder added increased. The silver powder content was dependent the amount of high-molecular-weight polymer resin, which enabled the silver powder to reach an appropriate viscosity. For example, the dry paste (>90 wt% silver powder) did not pass through the screen easily, so the screen was susceptible to rapid wear. If the paste was too wet (<75 wt% silver powder), the pattern was likely to collapse after screen printing; thus, the efficiency would be reduced. The optimal composition was 80~85 wt%. Second, the reported resistance data in the current literature were compared on the basis of the silver powder size, resistivity, and sintering temperature, as shown in Table 2. The universal sintering temperature is 800~900°C. According to Table 2, the conductivity declined when the particle size fell outside a range of 0.1~10 μm. The sheet resistance measured in this experiment was lower than other literature values, indicating that silver powder with a size of 1~1.5 μm sintered at 830~890°C is acceptable. In addition, when the silver powder particle size exceeded 10 μm, the powder was likely to stick on the screen. Most screens are 325 mesh, with an allowable range of approximately 10 μm. When the silver powder particle size is less than 0.1 μm, mass production is not feasible due to the high cost and burden of other additives; therefore, the particle size in most commercial silver pastes is approximately 1 μm, which is optimal. The third factor is the shape, and most commercial silver pastes use spherical silver powder. This experiment used spherical and angular silver powders to evaluate their differences. Some reported studies mixed multiangular or tree-like silver powder with spherical powder to increase conductivity [19], and others added a glass coating onto silver particles [19-21]; however, there was no absolute difference in the yield rates. In the present study, a paste was prepared using angular and spherical silver powders, and the difference in the sheet resistance was measured after sintering (4~9 mΩ/sq for the spherical powder and 7~13 mΩ/sq for the angular powder). Based on these results, the treatment of the silver powder is very important. Fourth, the silver paste was influenced by the atmosphere during sintering, and the ultimate objective in preparing the electrode was to maintain the original conductivity and prevent a severe decline in the electrical properties due to oxidation. Some studies have discussed whether silver powder is oxidized in the air or in glass [14, 22, 23]. Based on comparisons and discussions in multiple studies, a more specific reaction mechanism is proposed in this paper. As mentioned in [24], $Si+2AgO_{(in\ glass)} \rightarrow SiO_2+2Ag$, which indicates that the addition of $Bi_2O_3$ glass is effective for etching the antireflection layer of Si and SiNx. This experiment also used TEM to prove the formation of silicon oxide at the interface. Other studies have proposed a schematic of sintered silver crystallites on a silicon wafer. This study slightly corrected the model view proposed in Kontermann and Hörteis [25], as shown in Table 2.

Silver crystallites can grow increasingly large on the surface of the silicon wafer as the temperature increases, and their shape is no longer spherical but multiangular. The contact resistance of these individual silver crystallites has attracted the attention of scholars. Table 3 shows the data from the literature. The obtained $A_{crystalline\ silver,\ direct}$ values are large and closely related to the high sintering temperature, and the silver crystallites occupy a large proportion of the surface area of the silicon wafer. However, it is difficult to predict how many crystallites contact the silver layer in practice. The present literature indicates 0.1%~100%, and the value was 0.1% in this study, which has been reported in many other publications as well. Schubert and Huster [17] also calculated a contact resistance of $2\times10^{-8}$ $\Omega.cm^2$, which is close to the calculated value herein ($2.2\times10^{-8}$~$7.3\times10^{-8}$ $\Omega.cm^2$). Therefore, it was concluded that individual silver crystallites can reach a value of $10^{-8}$ $\Omega.cm^2$.

Table 1 Resistance comparison after firing at 830℃, 860℃, and 890℃.

| Firing temperature | $R_c$ ($\Omega$) | $\rho_C$ ($\Omega cm^2$) | $A_{crystalline\ silver}$ (%) | $A_{crystalline\ silver,\ direct}$ (%) | $\rho_{c,cryst}$ ($\Omega.cm^2$) |
|---|---|---|---|---|---|
| 830℃ | $4.09\times10^{-3}$ | $1.918\times10^{-4}$ | 15 | 0.1 | $2.9\times10^{-8}$ |
| 860℃ | $4.07\times10^{-3}$ | $2.922\times10^{-4}$ | 25 | 0.1 | $7.3\times10^{-8}$ |
| 890℃ | $4.31\times10^{-3}$ | $6.322\times10^{-4}$ | 35 | 0.1 | $2.2\times10^{-7}$ |

Table 2 Comparison of different conditions used for silver powders.

| Size | Resistivity | Firing temperature | Reference |
|---|---|---|---|
| 0.1 μm | 7 m$\Omega.cm^2$ | 750℃~840℃ | [32] |
| 1 μm | 1.7 m$\Omega.cm^2$ | 750℃~840℃ | [32] |
| 10 μm | 0.24 m$\Omega.cm^2$ | 750℃~840℃ | [32] |
| 50 nm -100 nm | 4.11 μ.cm | 450℃ | [33] |
| 1 μm -2 μm | 6.196 m$\Omega$/sq | 850℃ | [34] |
| 100 nm - 200 nm | 4.6 m$\Omega$/sq | 850℃ | [34] |
| 1.6 μm +20 nm | 4~11 μ.cm | 400℃~550℃ | [29] |
| 1.26 μm | 7~13 m$\Omega$/sq | 830℃~890℃ | System 1 |
| 1.56 μm | 4~9 m$\Omega$/sq | 830℃~890℃ | System 2 |

Table 3 Comparison of different parameters in the calculation of the specific contact resistance.

| $\rho_c$ ($\Omega \cdot cm^2$) | A crystalline silver (%) | A crystalline silver, direct (%) | $\rho_{c,crystallites}$ ($\Omega \cdot cm^2$) | Reference |
|---|---|---|---|---|
| $1.6 \times 10^{-3}$ | 10 | 0.1 | $1.6 \times 10^{-7}$ | [35, 36] |
| $0.094 \times 10^{-3}$ | 6.7 | 100 | $6.3 \times 10^{-6}$ | [37] |
| $0.047 \times 10^{-3}$ | 23.5 | 100 | $2 \times 10^{-4}$ | [37] |
| $4.4 \times 10^{-3}$ | 30.0 | 0.003 | $4 \times 10^{-8}$ | [17] |
| $22.2 \times 10^{-3}$ | 30.0 | 0.003 | $2 \times 10^{-8}$ | [17] |
| $4.9 \times 10^{-3}$ | 10.0 | 0.1 | $4.9 \times 10^{-7}$ | [38] |
| $1.918 \times 10^{-4}$ | 15 | 0.1 | $2.9 \times 10^{-8}$ | System 1 (830°C) |
| $2.922 \times 10^{-4}$ | 25 | 0.1 | $7.3 \times 10^{-8}$ | System 1 (860°C) |
| $6.322 \times 10^{-4}$ | 35 | 0.1 | $2.2 \times 10^{-7}$ | System 1 (890°C) |

Studies have examined the influence of the glass $T_g$ on silver paste [26-28] and described the relationship between the glass layer interfaces based on the $T_g$. DTA and SEM were used in this experiment to evaluate the influence of the $T_g$ on silver paste, and the $T_g$ controlled the melting of silver into the glass, influencing the precipitation of silver crystallites. The model of the $T_g$ is proposed in Fig. 5 A. From the figure, the higher the $T_g$ is, the thinner the glass layer, and the lower the coverage of the silver crystallites. On the other hand, the lower the $T_g$ is, the thicker the glass layer, and the greater the amount of melted silver in the glass. The glass powder developed by this experiment can uniformly precipitate crystalline silver and achieve the wetting of the silver layer on the silicon wafer. In addition to the $T_g$, the amount of added glass powder is also important (Fig. 5 B). Another study compared the addition of 3 wt% glass powder and 5 wt% glass powder to silver paste and proved that the thickness increased with the addition of glass powder. However, a high $T_g$ and small glass powder particle size are disadvantageous for commercialization [29]. Furthermore, most studies in the literature did not measure the thickness of the glass layer. In this study, the thickest part of the glass layer was measured at 2~4 μm, and the minimum thickness was calculated to be 0.2~0.8 μm using EPMA, EDS, and SIMS. The proposed model is shown in Fig. 5 C.

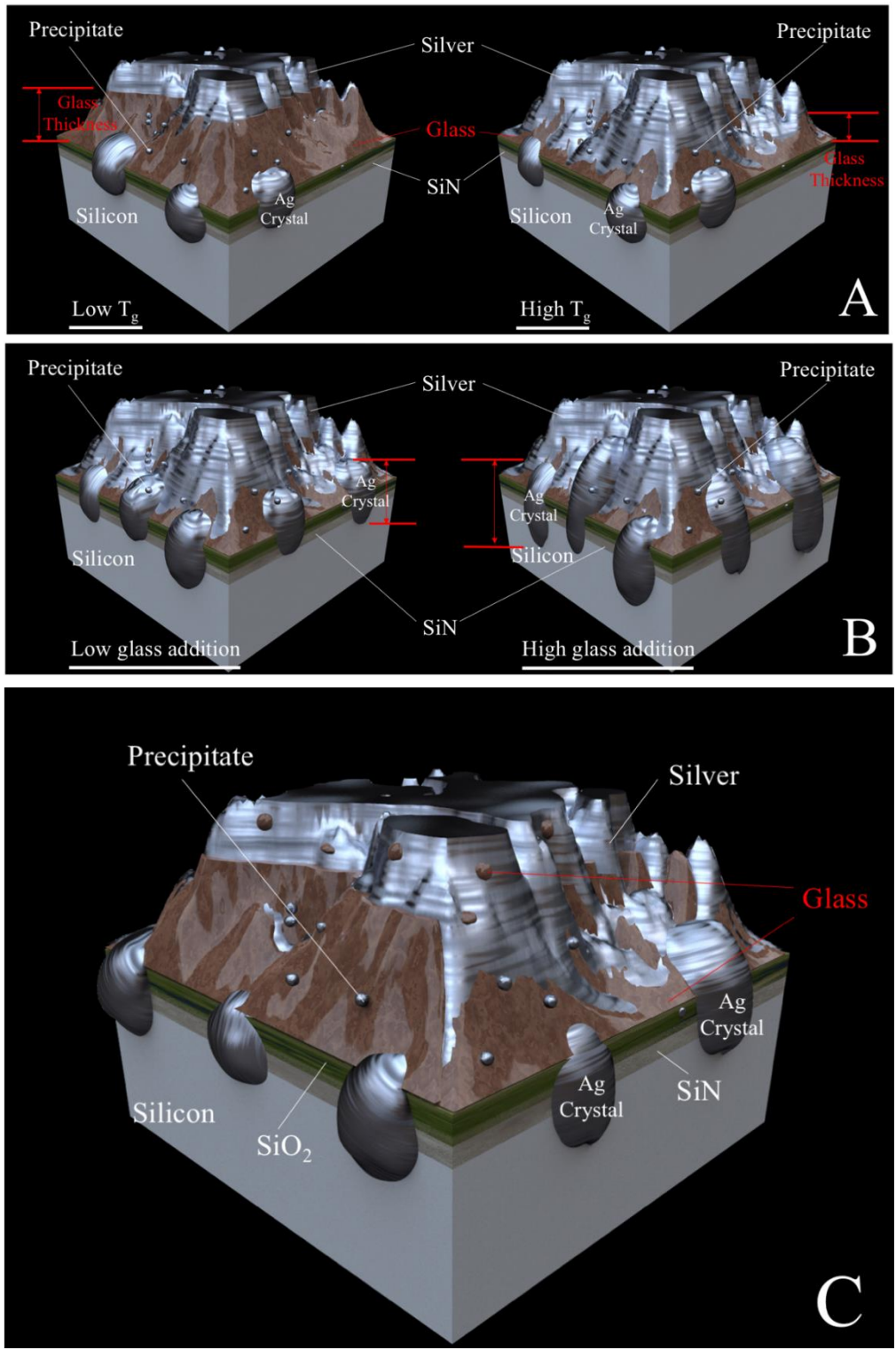

Fig. 5 (A) Model of a thick film at different glass transition temperatures. (B) Models of thick films with different glass additives. (C) Model of a TF contact after firing.

According to the DTA data in Fig. 6, there are three endothermic peaks (indicated by black arrowheads) in the CG powder at 225℃, 390℃, and 525℃. The Tg of homemade glass is 220℃, followed by a very wide exothermic peak, and then endothermic peaks occur at 495℃ and 625℃.

Fig. 7 A shows the surface of sintered Ag. The glass phase gradually precipitated, and a wetting effect at the grain boundary occurred; thus, glass densified the Ag during sintering. However, the precipitated glass phase between the grains was too wide, and the growth of Ag grains was obstructed. Fig. 8 A (a) shows the wetting effect of the glass phase, while Fig. 7 A (b) shows that the glass phase formed a continuous film with the silver grains and that the glass was somewhat soluble in the silver powder. The flow of the glass and silver powder particles moved; thus, the sintered volume changed rapidly, and Ag grains were further densified and coarsened, which minimized the duration of the sintering. From the figure, the Ag grains gradually increased in size with increasing temperature.

In addition, the glass powder particle size influenced the quality of the conducting silver paste. Kim and Lim [39] indicated that nearly all glass powder particle sizes were equal to or greater than the silver powder particle size when they were added to the conducting silver paste. However, this experiment showed that the glass powder particle size must match the silver powder particle size to obtain an optimal result. When the glass powder particle size was greater than that of the silver powder, the glass was likely to melt into the silver; thus, the glass phase, which was supposed to wet the silver, disappeared, and pores formed. In contrast, when the silver powder was larger in size than the glass powder, there was wetting of the glass phase in contact with the silver, and if a pore formed, the glass was likely to eliminate the pores. The melting and compactness can be expressed as follows:

$$S_R = S_B/S_A \qquad (2)$$

where $S_B$ is the silver solubility of the base, $S_A$ is the solubility of the glass additive, and $S_R$ is the ratio of the two solubilities.

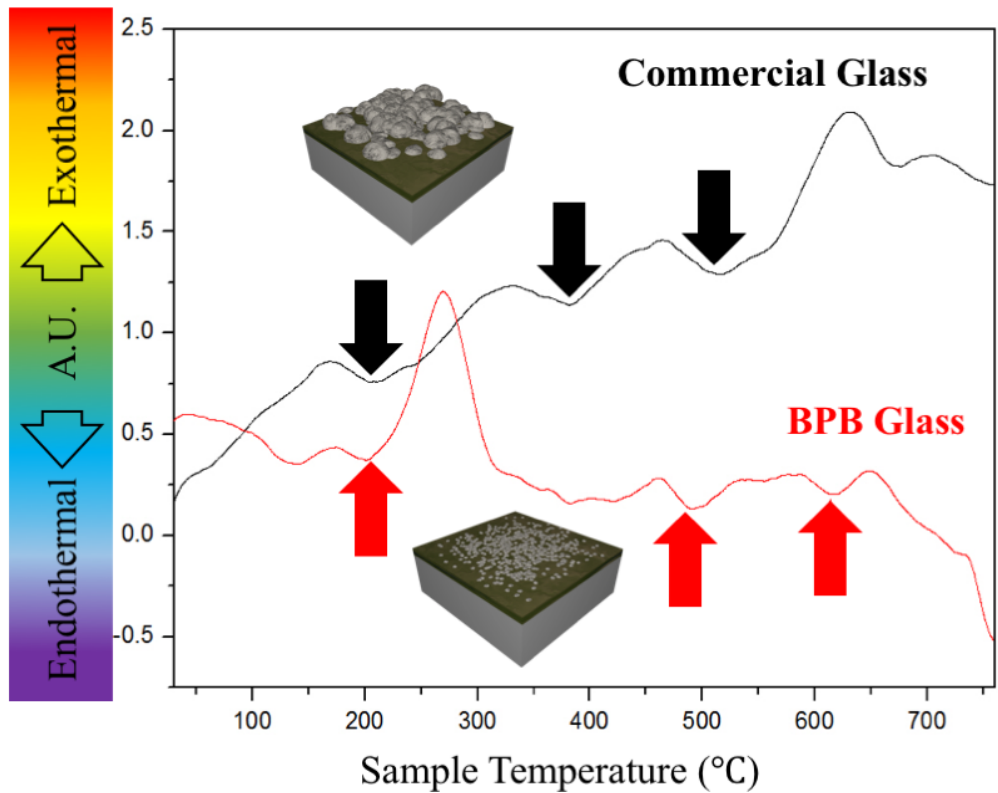

Fig. 6 DTA curves of the glass frits for commercial glass (CG) and homemade glass (BPB).

If $S_B/S_A<1$, according to Fig. 7 A (a) (red arrow), the glass phase thickens as the temperature increases, pores form in the glass and increase in size, and the glass melts into the silver powder. When $S_B/S_A>1$, according to Fig. 7 A (b) (red arrow), the glass phase flows into the gaps between the silver particles as the temperature increases and successfully wets the silver grains, and the glass phase on the surface is narrowed.

Fig. 7 B shows the cross-section of an Ag film on a silicon substrate. The approximate positions of the glass layer, Ag layer and silicon substrate were identified by EDS; the glass layer thickness increased with temperature. In Fig. 7 B (a), the Ag crystallites precipitated in the glass layer, while some Ag precipitated on the silicon wafer. There were additional Ag crystallites in the glass layer, which were continuously distributed and spherical in shape. In Fig. 7 B (b), spherical Ag precipitates were not continuously distributed along the interface in the glass layer. The Ag crystallite particles gradually increased in size, changed from spherical to angular in shape and increased in their contact with the silicon substrate as the temperature increased.

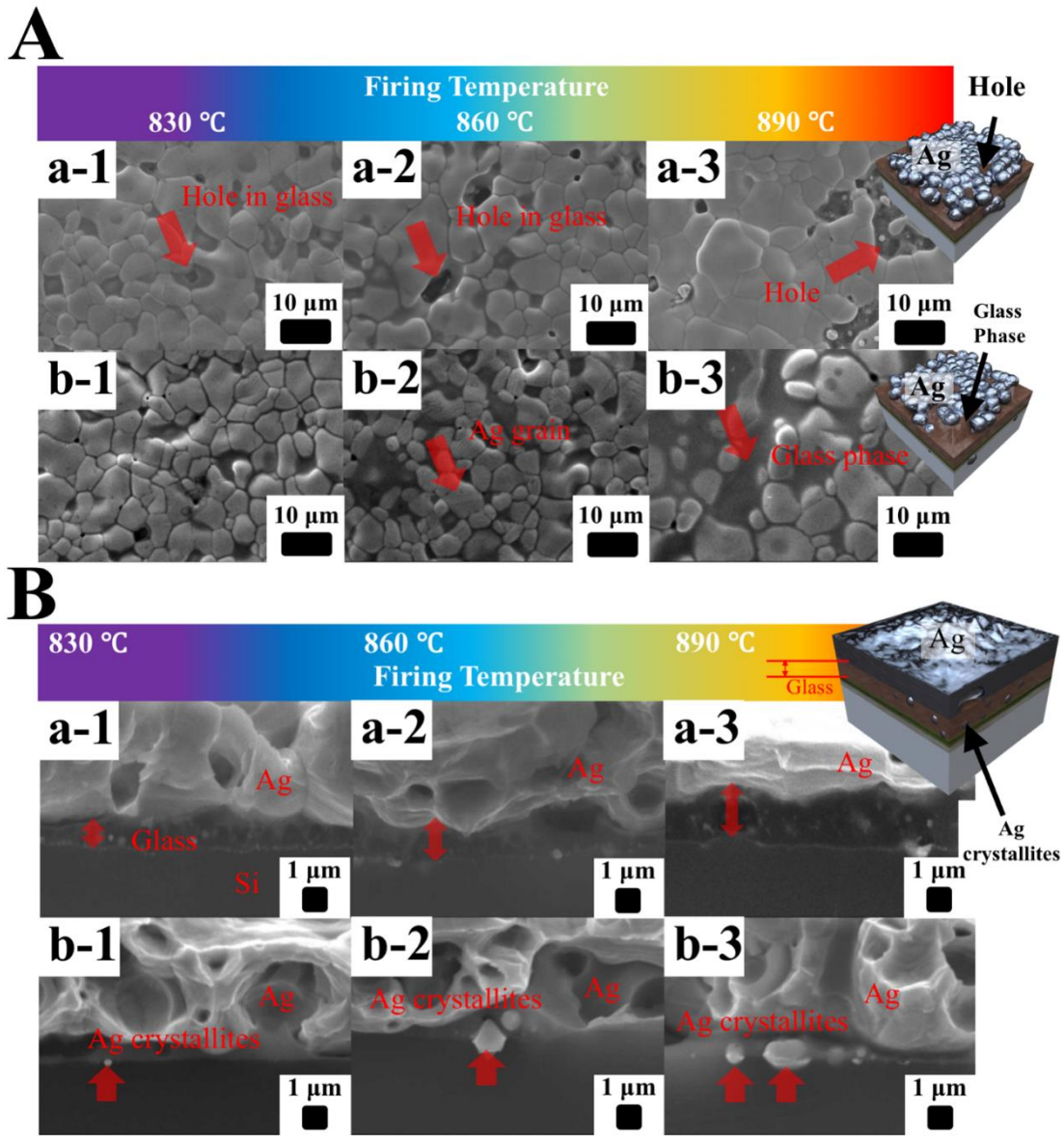

Fig. 7 (A) SEM micrographs of surfaces of (a) CG and (b) homemade glass mixed with silver paste. (B) Cross-sections of (a) CG and (b) homemade glass mixed with silver paste.

The literature indicates that the optimal addition of glass powder is 3~5 wt% [21] with a silver powder content of 80~85 wt%; thus, this experiment used homemade glass powder to prove these findings. The $T_g$ of the added glass phase was lower than that found in the literature, and the glass powder particle size, D50, was 0.8 μm~1.2 μm. Glass powder was added at 2 wt% and 4 wt%, and firing occurred at 830℃, 860℃, and 890℃, as shown in Fig. 8 B.

To study the condition of the interface after the glass powder was added to the silver paste and silicon during sintering, a fractured sample section was polished and then surface scanned by EPMA to analyze the element distribution. Silver, oxygen, bismuth, silicon, and phosphor were detected. The top layer was silver paste, the lower layer was crystalline silicon, and the middle layer was glass. These samples were numbered #1 and #9 in system 3. Fig. 8 A shows the results for 2 wt% glass powder (sample #1 in system

3) fired at 830℃, and the precipitation of bismuth with oxygen can be observed. When bismuth glass powder was added, the bismuth glass flowed to the bottom as the sintering temperature increased; thus, the levels of bismuth and oxygen were interdependent. The bismuth glass was between the silicon layer and silver layer; however, the glass layer was not continuously distributed. The thickness of the glass layer was 2~3 μm, which may have been a result of the 830℃ sintering temperature being slightly lower than the 840℃ eutectic temperature for silicon and silver. In Fig. 8 B, with the addition of 4 wt% glass powder (sample #9 in system 3) and firing at 890℃, the top silver paste glass frit gradually diminished, and most of the frit flowed to the bottom layer and bonded with the crystalline silicon. Comparison of the glass layer after firing at 890℃ and 830℃ revealed that the thickness of the glass layer increased with the temperature. According to the DTA data in Fig. 6 of the previous section, the glass softening point was 220℃; therefore, the bismuth glass had already flowed into this interface and formed a glass layer before the silver bonded with the silicon. When the amount of glass frit in the top layer of the silver increased, their distributions were not uniform; it seems the glass did not melt into the silver paste but agglomerated. In contrast, the addition of glass powder resulted in a decreased amount of glass frit in the top layer.

SIMS was used to analyze the surface elements and glass layer thickness, which cannot be observed by EPMA but were indirectly obtained using this method. However, the part of the silver that was not covered with a glass layer was corroded, and the actual thickness of the glass layer could not be determined; however, a reference value could be used. Different combinations of the glass additive content and sintering temperature were compared. As shown in Fig. 8 C (a), upon adding 4 wt% glass and firing at 830℃, the intensity of the bismuth signal exceeded that of the silver signal at 0.36 μm. Fig. 8 C (b) shows the results for 2 wt% glass and sintering at 860℃, where the bismuth signal intensity exceeded the silver signal intensity at 0.65 μm. Fig. 8 C (c) shows the results for 2 wt% glass and sintering at 890℃, where the bismuth signal intensity exceeded the silver signal intensity at 0.76 μm. Fig. 8 Ct presents the results of the peak attenuation with depth. Comparison of all three parameters shows that a small amount of added glass powder can significantly change the sintering of silver paste on silicon wafers. When the bismuth was analyzed, the intensity of the silver signal gradually approached that of the bismuth signal as the analysis depth increased, and the intensity of the silver signal exceeded that of the bismuth signal for the first time at 0.3~0.8 μm. Therefore, when 2 wt% or 4 wt% glass powder was added to the glass layer with a firing temperature of 830~890℃, a 0.2~0.8 μm glass layer formed at the interface.

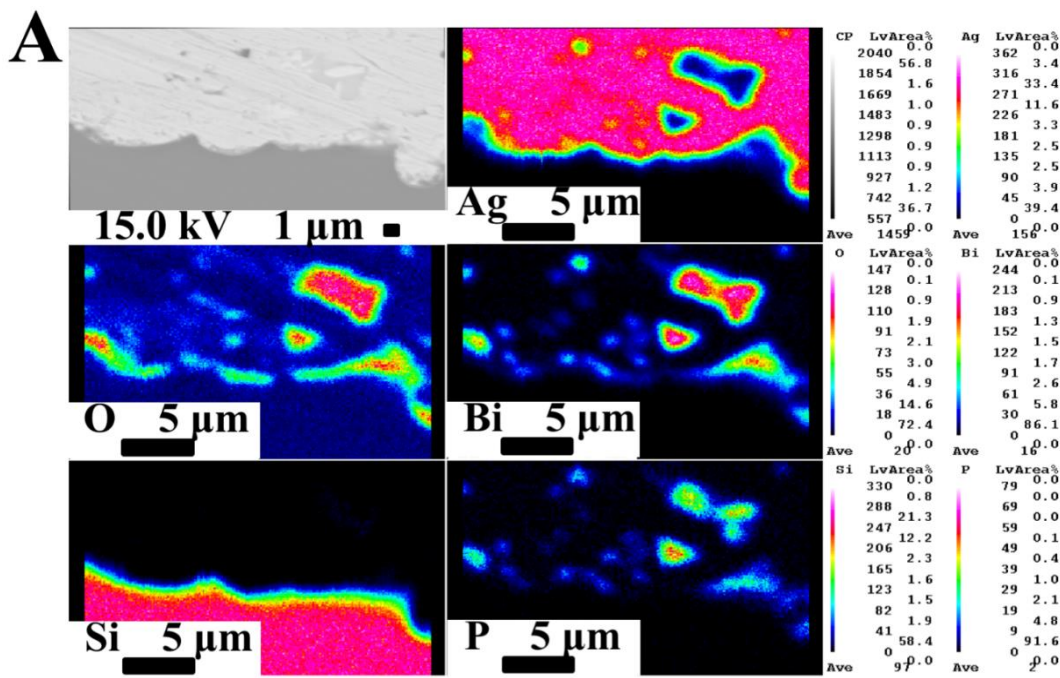
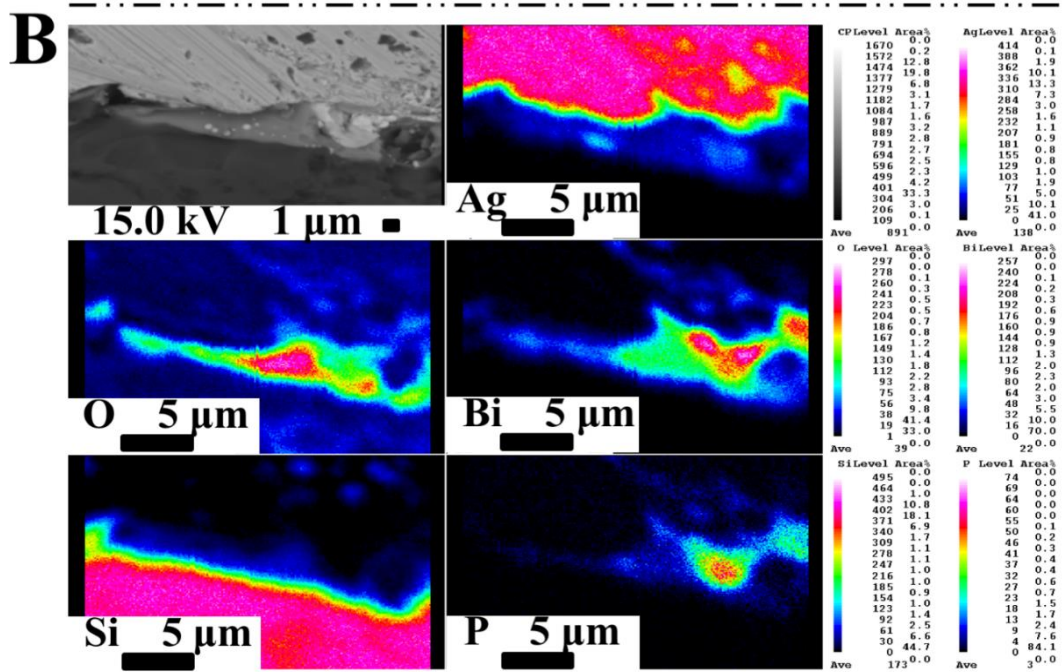

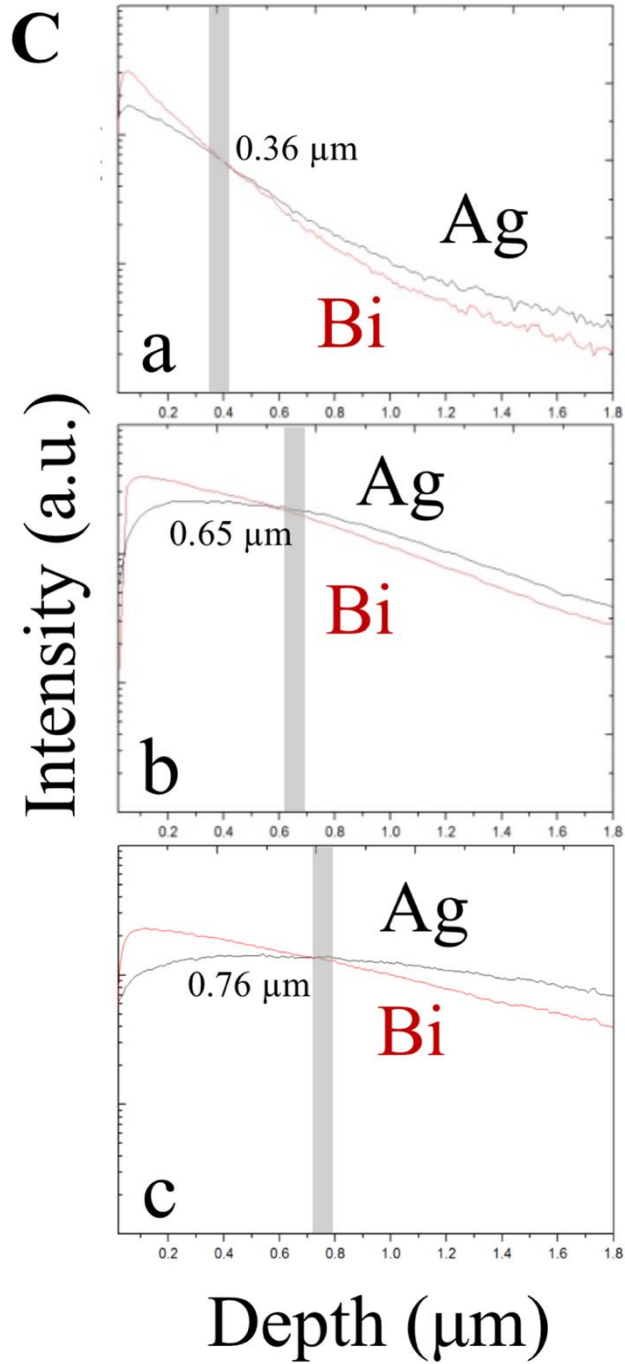

Fig. 8 Dot mapping of Ag, O, Bi, Si, and P for (A) specimen 1 in system 3 (processed at 830℃ for 10 min) and (B) specimen 9 in system 3 (processed at 890℃ for 10 min). (C) SIMS results for etched surfaces of the (a) #3, (b) #4 and (c) #7 silver paste specimens in system 3.

After analyzing the surface elements and the glass layer thickness, TEM was used to carefully observe the interface, and different crystalline phases were analyzed by EDS. The results are shown in Fig. 9 A. Qualitative analysis was completed for three different crystallization points at the interface. This region was in the conversion zone, and the composition could not be accurately analyzed due to limitations of the EDS technique; however, the element contents provided some clues. Fig. 9 A (a) shows the precipitation of silver in the glass phase; the detected elements included silver, bismuth, oxygen, and silicon. Fig. 9 A (b) shows the precipitates on the silicon interface. The silver signal was very strong, although in the silicon area, the precipitate particles were concentrated within 100 μm of the interface, and the particles were nearly spherical nanoscale silver. Fig. 9 A (c) shows that the silicon signal was detected deep in the silicon area, whereas the silver signal was suppressed in that area. Fig. 9 A (a)-(c) proves that the silver diffused into the silicon. Then, the method proposed by Rollert and Stolwijk [38] was used to calculate the diffusivity and activation energy.

Ag crystallites in the glass layer were observed by high-resolution TEM (HRTEM). The lattice distance was 0.238 nm, as measured in Fig. 9 B. The value is quite close to the lattice spacing of 0.236 μm for (111) Ag planes, meaning that the Ag crystallites did not react with the glass phase during nucleation and growth. Herein, the silver was oxidized through gas phase diffusion when it melted into the glass during sintering, and then the Ag nanoparticles precipitated during cooling.

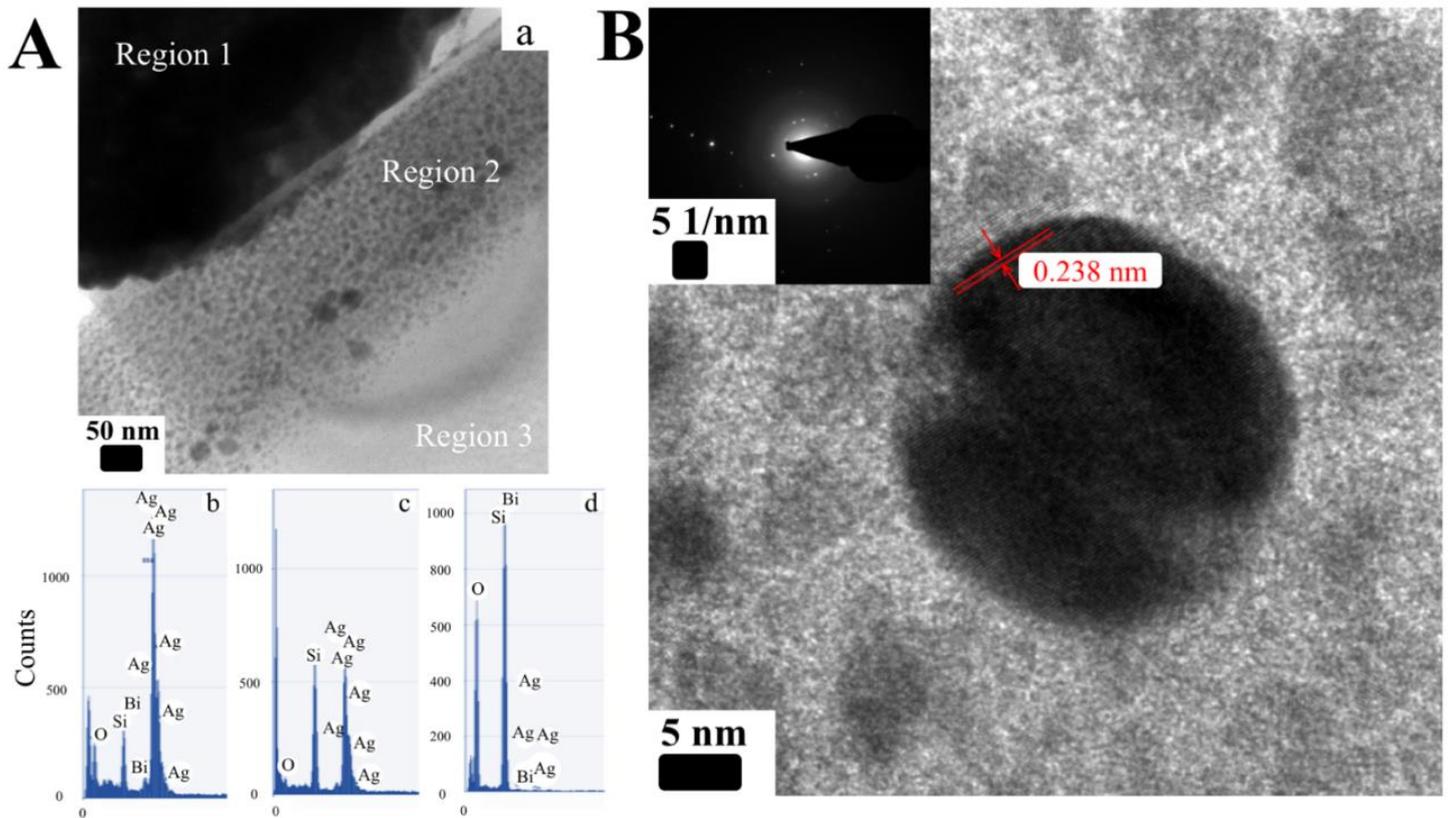

Fig. 9 (A) (a) TEM bright-field images of specimen #2 and EDS analysis of (b) region 1, (c) region 2 and (d) region 3 for silver paste system 5. (B) High-resolution TEM image of silver crystallites in the glass layer.

**Conclusion**

In this study, we successfully fabricated a better metal-glass composite material that has superior mechanical and electrical properties, especially for ohmic contacts in ICs and solar cells, by a screen printing technique. During this study, the properties of the metal composite were investigated in depth. The structural properties of screen-printed Ag/Si contacts were investigated using SEM, EPMA and SIMS. The specific contact resistance was determined using the TLM pattern. The corroded surface was used to study the microstructures and features at the contact interfaces.

Specifically, the Ag layer contacted the interface of the silicon substrate by precipitating nanoscale Ag crystallites on the interface to allow the electric current to smoothly flow through the substrate and reach the Ag layer. In fact, the Ag crystallites etched into the silicon substrate. The higher the temperature was, the more obvious the etching phenomenon. The following layers of the microstructure were observed: (1) silver crystallites connected to the silver directly; (2) silver crystallites in the glass; (3) glass; and (4) spherical silver nanoparticles in the glass layer. The preferred orientation of the silver surface was (200), and the silver crystallites that melted into the glass layer had a (111) preferred orientation. The minimum measured contact resistance was $1.92 \times 10^{-4}$ ($\Omega.cm^2$), the maximum was $6.3 \times 10^{-4}$ ($\Omega.cm^2$), the minimum contact resistance of a single silver crystal was $2.9 \times 10^{-8}$ ($\Omega.cm^2$), and the maximum was $2.2 \times 10^{-7}$ ($\Omega.cm^2$). It is concluded that the resistivity of the silver crystallites increased due to the influence of the temperature.

Moreover, the thickness of the interface glass layer varied with the $Bi_2O_3$ glass content. At low sintering temperatures, the thickness of the interface glass layers was low, and few Ag crystallites covered the interface. At high sintering temperatures, the thickness of the interface glass layers was high, and many Ag crystallites were present in the glass. The glass powder developed herein can uniformly precipitate Ag crystallites and achieve wetting of the silver surface and silicon wafer. Although the $T_g$ of both glass powders was very low, the homemade glass powder produced an intense exothermic peak at 275℃, indicating that a crystalline phase had formed, which obstructed the silver from melting directly into the glass. In contrast, the commercial glass powder did not produce an intense exothermic peak, causing a large amount of silver to melt into the glass phase until it became saturated and caused the glass to flow to the interface, resulting in pores on the silver surface. The SEM, EPMA, SIMS and TEM results also confirmed that the precipitation of Ag crystallites was followed by the formation of an interface glass layer, and the continuity of the interface glass layer was controlled by the amount of added glass.

This study provides a detailed analysis of well-controlled electrical properties, morphologies, and experimental details for silver contacts made from a novel composition consisting of $Bi_2O_3$, $B_2O_3$, and additional $P_2O_5$ mixed with Ag, which created properties and stabilities different from those obtained using a general formula consisting of commercial $Bi_2O_3$ glass powder ($Bi_2O_3$ and $B_2O_3$) mixed with Ag. This in-depth study provides more options and experimental details for metal contact processing for existing applications in TF electronics, solar cells, sensor cells, and ICs.

**References**

1. Lin, W.P., et al. *Microstructures of silver films plated on different substrates and annealed at different conditions*. in *Electronic Components and Technology Conference (ECTC), 2011 IEEE 61st*. 2011.

2. Zhang, C. and C.W. Bates Jr, *Metal-mediated crystallization in Si–Ag systems.* Thin Solid Films, 2009. **517**(19): p. 5783-5785.

3. Ko, Y.N., et al., *Characteristics of Pb-based glass frit prepared by spray pyrolysis as the inorganic binder of silver electrode for Si solar cells.* Journal of Alloys and Compounds, 2010. **490**(1-2): p. 582-588.

4. Yi, J.H., et al., *Fine size Pb-based glass frit with spherical shape as the inorganic binder of Al electrode for Si solar cells.* Journal of Alloys and Compounds, 2010. **490**(1-2): p. 488-492.

5. Jeon, S.J., S.M. Koo, and S.A. Hwang, *Optimization of lead- and cadmium-free front contact silver paste formulation to achieve high fill factors for industrial screen-printed Si solar cells.* Solar Energy Materials and Solar Cells, 2009. **93**(6-7): p. 1103-1109.

6. Tsai, J.-T. and S.-T. Lin, *Silver powder effectiveness and mechanism of silver paste on silicon solar cells.* Journal of Alloys and Compounds, 2013. **548**(0): p. 105-109.

7. Hilali, M.M., et al., *Understanding the formation and temperature dependence of thick-film Ag contacts on high-sheet-resistance Si emitters for solar cells.* Journal of the Electrochemical Society, 2005. **152**(Compendex): p. G742-G749.

8. Hilali, M., et al. *Optimization of self-doping Ag paste firing to achieve high fill factors on screen-printed silicon solar cells with a 100 /sq. emitter*. in *29th IEEE Photovoltaic Specialists Conference, May 19, 2002 - May 24, 2002*. 2002. New Orleans, LA, United states: Institute of Electrical and Electronics Engineers Inc.

9. Kim, J.H., et al., *Characteristics of Bi-based glass frit having similar mean size and morphology to those of silver powders at high firing temperatures.* Journal of Alloys and Compounds, 2010. **497**(1-2): p. 259-266.

10. Satpati, B., et al., *Nanoscale ion-beam mixing in Au–Si and Ag–Si eutectic systems.* Applied Physics A: Materials Science & Processing, 2004. **79**(3): p. 447-451.

11. Zhang, X.N., Z. Zhang, and C.R. Li, *Growth of one-dimensional Ag/Si/SiO$_x$ capsule nanostructures by self-assembled SiO$_x$ template.* Applied Physics A: Materials Science & Processing, 2005. **81**(1): p. 163-167.

12. Porter, L.M., A. Teicher, and D.L. Meier, *Phosphorus-doped, silver-based pastes for self-doping ohmic contacts for crystalline silicon solar cells.* Solar Energy Materials and Solar Cells, 2002. **73**(2): p. 209-219.

13. Karski, S., et al., *Interaction between Pd and Ag on the surface of silica.* Journal of Molecular

Catalysis A: Chemical, 2005. **240**(1-2): p. 155-163.

14. Assal, J., B. Hallstedt, and L.J. Gauckler, *Experimental phase diagram study and thermodynamic optimization of the Ag-Bi-O system.* Journal of the American Ceramic Society, 1999. **82**(3): p. 711-715.

15. Hu, M., et al., *The effects of nanoscaled amorphous Si and SiNx protective layers on the atomic oxygen resistant and tribological properties of Ag film.* Applied Surface Science, 2012. **258**(15): p. 5683-5688.

16. Iida, S., et al., *Thin Ag film formation onto Si/SiO2 substrate.* Applied Surface Science, 2000. **166**(1–4): p. 160-164.

17. Schubert, G., F. Huster, and P. Fath, *Physical understanding of printed thick-film front contacts of crystalline Si solar cells--Review of existing models and recent developments.* Solar Energy Materials and Solar Cells, 2006. **90**(18-19): p. 3399-3406.

18. Kontermann, S., R. Preu, and G. Willeke, *Calculating the specific contact resistance from the nanostructure at the interface of silver thick film contacts on n-type silicon.* Applied Physics Letters, 2011. **99**(11).

19. Ko, Y.N., et al., *Characteristics of silver-glass composite powders as the silver electrode for Si solar cells.* Journal of Alloys and Compounds, 2010. **491**(1-2): p. 584-588.

20. Yi, J.H., et al., *Characteristics of Ag powders coated with Pb-based glass material prepared by spray pyrolysis under various gas environments.* Ceramics International, 2010. **36**(8): p. 2477-2483.

21. Koo, H.Y., et al., *Nano-sized silver powders coated with Pb-based glass material with high glass transition temperature.* Colloids and Surfaces A: Physicochemical and Engineering Aspects, 2010. **361**(1-3): p. 45-50.

22. Tang, K., et al., *Thermochemical and Kinetic Databases for the Solar Cell Silicon Materials Crystal Growth of Si for Solar Cells*, K. Nakajima and N. Usami, Editors. 2009, Springer Berlin Heidelberg. p. 219-251.

23. Assal, J., B. Hallstedt, and L.J. Gauckler, *Thermodynamic Assessment of the Silver–Oxygen System.* Journal of the American Ceramic Society, 1997. **80**(12): p. 3054-3060.

24. Huh, J.-Y., et al., *Effect of oxygen partial pressure on Ag crystallite formation at screen-printed Pb-free Ag contacts of Si solar cells.* Materials Chemistry and Physics, (0).

25. Kontermann, S., et al., *Physical understanding of the behavior of silver thick-film contacts on n-type silicon under annealing conditions.* Solar Energy Materials and Solar Cells, 2009. **93**(9): p. 1630-1635.

26. Koo, H.Y., J.H. Yi, and Y.C. Kang, *Characteristics of Bi-based glass powders with various glass transition temperatures prepared by spray pyrolysis.* Ceramics International, 2010. **36**(5): p. 1749-1753.

27. Zhang, Y., et al., *Thermal properties of glass frit and effects on Si solar cells.* Materials Chemistry and Physics, 2009. **114**(1): p. 319-322.

28. Shim, S.-B., et al., *Wetting and surface tension of bismate glass melt.* Thermochimica Acta, 2009. **496**(1-2): p. 93-96.

29. Park, S., D. Seo, and J. Lee, *Preparation of Pb-free silver paste containing nanoparticles.* Colloids

and Surfaces A: Physicochemical and Engineering Aspects, 2008. **313-314**(Compendex): p. 197-201.

30. Jean, J.-H. and C.-R. Chang, *Interfacial reaction kinetics between silver and ceramic-filled glass substrate.* Journal of the American Ceramic Society, 2004. **87**(7): p. 1287-1293.

31. Weber, L., *Equilibrium solid solubility of silicon in silver.* Metallurgical and Materials Transactions A: Physical Metallurgy and Materials Science, 2002. **33**(4): p. 1145-1150.

32. Hilali, M.M., et al., *Effect of Ag particle size in thick-film Ag paste on the electrical and physical properties of screen printed contacts and silicon solar cells.* Journal of the Electrochemical Society, 2006. **153**(Compendex): p. A5-A11.

33. Park, K., D. Seo, and J. Lee, *Conductivity of silver paste prepared from nanoparticles.* Colloids and Surfaces A: Physicochemical and Engineering Aspects, 2008. **313-314**(Compendex): p. 351-354.

34. Rane, S.B., et al., *Firing and processing effects on microstructure of fritted silver thick film electrode materials for solar cells.* Materials Chemistry and Physics, 2003. **82**(1): p. 237-245.

35. Ballif, C., et al., *Silver thick-film contacts on highly doped n-type silicon emitters: Structural and electronic properties of the interface.* Applied Physics Letters, 2003. **82**(12): p. 1878-1880.

36. Ballif, C., et al. *Nature of the Ag-Si interface in screen-printed contacts: a detailed transmission electron microscopy study of cross-sectional structures*. in *Photovoltaic Specialists Conference, 2002. Conference Record of the Twenty-Ninth IEEE*. 2002.

37. Cabrera, E., et al. *Current transport in thick film Ag metallization: Direct contacts at Silicon pyramid tips?* in *1st International Conference on Crystalline Silicon Photovoltaics, SiliconPV 2011, April 17, 2011 - April 20, 2011*. 2011. Freiburg, Germany: Elsevier Ltd.

38. Kontermann, S., G. Willeke, and J. Bauer, *Electronic properties of nanoscale silver crystals at the interface of silver thick film contacts on n-type silicon.* Applied Physics Letters, 2010. **97**(19): p. 191910.

39. Kim, B.S., et al., *Effect of Bi2O3 content on sintering and crystallization behavior of low-temperature firing Bi2O3-B2O3-SiO2 glasses.* Journal of the European Ceramic Society, 2007. **27**(2-3): p. 819-824.

40. Maeder, T., *Review of Bi2O3 Based Glasses for Electronics and Related Applications*. International Materials Reviews 58.1 (2013): 3-40. Web.